\documentclass[english,aps,prb,twocolumn,showpacs]{revtex4-1}

\usepackage{graphics}
\usepackage{epsf}
\usepackage{epsfig}

\begin{document}

\title{Theory of magnetism in La$_2$NiMnO$_6$}
\author{Prabuddha Sanyal$^{1}$} 
\affiliation{$^{1}$ IIT Roorkee, Roorkee 247667, India}


\begin{abstract}
The magnetism of ordered and disordered La$_2$NiMnO$_6$ is explained using a model involving
double exchange and superexchange. 
 The concept of majority spin hybridization in the large
  coupling limit is used to explain the ferromagnetism of La$_2$NiMnO$_6$ as compared to
 the ferrimagnetism of Sr$_{2}$FeMoO$_{6}$.
 The ferromagnetic insulating ground state in the ordered phase is explained. The essential role played by the 
 Ni-Mn superexchange between the Ni $e_{g}$ electron spins and the Mn $t_{2g}$ core electron spins 
 in realizing this ground state, is outlined.
 In presence of antisite disorder, the model system is found to exhibit a tendency of
  becoming a spin-glass at low temperatures,
 while it continues to retain a ferromagnetic transition at higher temperatures, similar to recent
 experimental observations [D. Choudhury {\it .et.al.}, Phys. Rev. Lett. {\bf 108}, 127201 (2012)].
 This reentrant spin-glass or reentrant ferromagnetic behaviour is explained 
 in terms of the competition of the ferromagnetic double exchange
 between the Ni $e_{g}$ and the Mn $e_{g}$ electrons, and the ferromagnetic Ni-Mn superexchange, with the
 antiferromagnetic antisite Mn-Mn superexchange.  
\end{abstract}

\maketitle

\section{Introduction}

The double perovskite (DP) La$_2$NiMnO$_6$ (LNMO), has generated a lot of interest for its  
magnetodielectric properties~\cite{AdvMat}, making it a promising candidate
 for potential device applications~\cite{Guo}. There have also been   
  suggestions of topological phases~\cite{ArunParam} for LNMO formed in LaNiO$_{3}$-LaMnO$_{3}$ superlattices.
  The pure compound is deemed to be a ferromagnetic semiconductor~\cite{AdvMat,GoodenoughPRB},
  with a Curie temperature very close
 to room temperature (T$_{c}\approx$ 280K).  Recently, there has been reports
 of reentrant spin-glass behaviour in partially disordered LNMO at low temperatures,
 along with a disordered ferromagnetism at higher temperatures~\cite{debraj}. In this paper,
 a simple theoretical model for LNMO is proposed which can explain the ferromagnetic
 insulating behaviour of the ordered compound, as well as provide insight into the
 low temperature spin-glass behaviour observed in the disordered case. Since there is supposed to be 
 significant contribution of the relative spin-orientation dependent assymetric hopping between 
 the transition metal sites to the dielectric constant, hence the colossal magnetodielectricity
 is closely related to the magnetism~\cite{debraj}. Hence an understanding of the magnetic and electronic
 properties of this material is essential to the understanding of the magnetodielectricity in this material. 


\begin{figure}
\includegraphics[width=7cm,height=6cm]{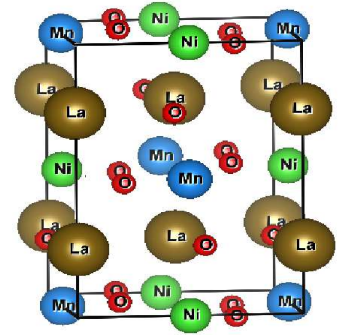} \\
\caption{(colour online) Crystal structure of La$_{2}$NiMnO$_{6}$ }
\label{structure}
\end{figure}

\begin{figure}
\includegraphics[width=8cm]{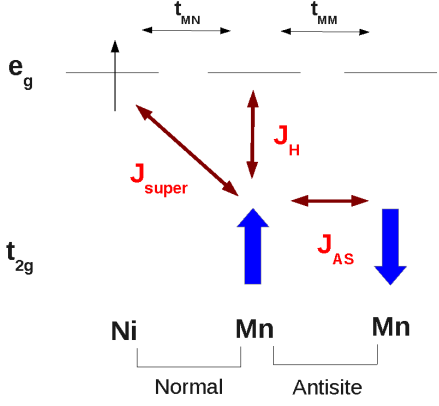} \\
\caption{(colour online) Exchange mechanisms for model Hamiltonian}
\label{exchangepath}
\end{figure}

\begin{figure}
\includegraphics[width=8cm]{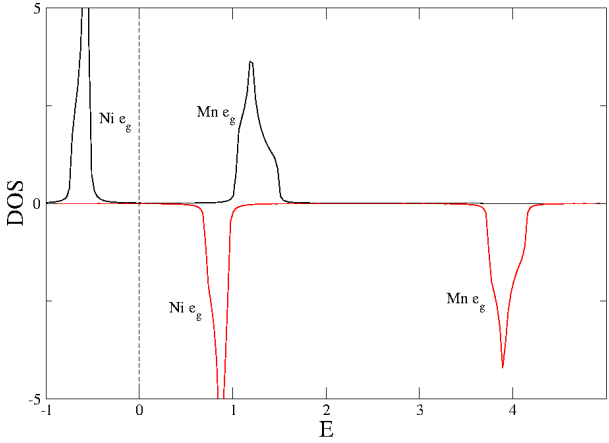} \\
\caption{(colour online) $e_{g}$ electron DOS for ordered La$_{2}$NiMnO$_{6}$, from model Hamiltonian. Dotted line: Fermi
 energy.}  
\label{DOS}
\end{figure}

\section{The model Hamiltonian}

In La$_{2}$NiMnO$_{6}$ the Nickel is in $Ni^{2+}$ state ($t_{2g}^{6}e_{g}^{2}$) and has two $e_{g}$ electrons, 
while the Manganese
 is in $Mn^{4+}$ state ($t_{2g}^{3}e_{g}^{0}$) and has three $t_{2g}$ electrons~\cite{AdvMat,debraj,Tanusri}.
  As the $t_{2g}$ electrons are more localized, and are
 parallel due to strong Hund coupling, they may be thought of as a core classical spin $S=3/2$. Nickel has
 a filled $t_{2g}$ shell, and net $t_{2g}$
 spin $S=0$, hence this compound can be thought of as a manganite where half the sites do not have a core spin.
  When Nickel $e_{g}$ electrons  hop on to the vacant $e_{g}$ orbitals of Manganese,
 they have an exchange with the 
 large Mn $t_{2g}$ core spins as usual, but the difference with most other DP-s
 like Sr$_{2}$FeMoO$_{6}$ (SFMO)~\cite{Millis} 
 and Sr$_{2}$CrOsO$_{6}$ (SCOO)~\cite{Sr2CrOsO6} is that this exchange coupling is
 ferromagnetic rather than antiferromagnetic.
  A model with somewhat similar ingredients had been proposed in Ref~\cite{Sanjeev},
  but they only considered the ordered case numerically, and at zero temperature.
  A detailed understanding of the magnetism of LNMO including the case of antisite disorder is lacking
  so far. We propose the following simple model Hamiltonian for the ordered case:
\begin{eqnarray}
H_{ord}&=&\epsilon_{Ni}\sum_{i\sigma}c_{N,i\sigma}^{\dagger}c_{N,i\sigma}+\epsilon_{Mn}\sum_{i\sigma}c_{M,i\sigma}^{\dagger}c_{M,i\sigma} \nonumber \\
 &+&t_{MN}\sum_{<ij>\sigma}c_{M,i\sigma}^{\dagger}c_{N,j\sigma}+J_{H}\sum_{i\alpha,\beta}c_{M,i\alpha}^{\dagger}\vec{\sigma}_{\alpha\beta}
c_{M,i\beta}\cdot \vec{S}_{i} \nonumber \\
 &+&J_{super}\sum_{<ij>} c_{N,j\alpha}^{\dagger}\vec{\sigma}_{\alpha\beta}c_{N,j\beta}\cdot \vec{S}_{i} 
\label{finiteJH}
\end{eqnarray}
where $c_{N,i\sigma}^{\dagger}$ ($c_{M,i\sigma}^{\dagger}$) creates an electron at the i-th Ni(Mn) site 
with spin $\sigma$. $\epsilon_{Ni}$ and $\epsilon_{Mn}$ are site energies of $e_{g}$ electrons 
at the Ni and Mn sites respectively, while $t_{MN}$ represents the hopping between the $e_{g}$ orbitals of Ni and Mn. 
In our attempt to find the simplest Hamiltonian which can explain the magnetism of LNMO,
 we consider a single orbital model~\cite{singleorb}. In addition to $J_{H}$ which results in Ni-Mn $e_{g}-e_{g}$ double exchange,
 this Hamiltonian introduces for LNMO the
 Ni-Mn $e_{g}-t_{2g}$ superexchange $J_{super}$ 
(see Fig~\ref{exchangepath}), 
which will be found to set the scale of the ferromagnetic
 $T_{c}$, similar to the Cr-Os superexchange in SCOO~\cite{Mohit}.
 Such a superexchange had earlier been calculated~\cite{Tanusri} for LNMO in the context of Kugel-Khomskii model
~\cite{Kugel-Khomskii} and found to be ferromagnetic~\cite{superJ1J2}. The parameter choice is phenomenological, 
but motivated from the DFT results of Ref~\cite{Tanusri}.
 It is to be noted that the Kondo coupling  between the 
core and itinerant spin on the B site is ferromagnetic for LNMO (and is equal to the Hund coupling $J_{H}$),
 and not antiferromagnetic as in case of SFMO~\cite{Millis,mePinaki} (See Appendix). If one considers
 the limit of large coupling, $J_{H}\rightarrow-\infty$~\cite{JHinfty,AndHas,DeGennes}, then this Hamiltonian simplifies further:
\begin{eqnarray}
H^{(1)}_{ord}&=&\epsilon_{Ni}\sum_{i\sigma}c_{N,i\sigma}^{\dagger}c_{N,i\sigma}+\tilde{\epsilon}_{Mn}\sum_{i}m_{i}^{\dagger}m_{i} \nonumber \\
 &+&t_{MN}\sum_{<ij>}(cos\frac{\theta_{i}}{2}m_{i}^{\dagger}c_{N,j\uparrow}+sin\frac{\theta_{i}}{2}e^{i\phi_{i}}m_{i}^{\dagger}c_{N,j\downarrow}) \nonumber \\
 &+&J_{super}\sum_{<ij>} c_{N,j\alpha}^{\dagger}\vec{\sigma}_{\alpha\beta}c_{N,j\beta}\cdot \vec{S}_{i} 
\label{Jinfty}
\end{eqnarray}
where $m^{\dagger}$ represents spinless Mn degree of freedom, $\theta_{i}$ is the polar angle between the spin $\vec{S}_{i}$ at i-th Mn site with z-axis, $\phi_{i}$ is the azimuthal angle, and
charge transfer energy is given by $\Delta=\tilde{\epsilon}_{Mn}-\epsilon_{Ni}$.
 This represents the minimal model for understanding
the magnetism of ordered LNMO. It is to be noted that this Hamiltonian has
{\em majority spin hybridization} between Mn and Ni, i.e., in case of a fully ferromagnetic arrangement 
 of the B-site (Mn) core spins ($\theta_{i}=0$, $\phi_{i}=0$ $\forall i$), {\em B-B$^{\prime}$ (Mn-Ni) hybridization 
  in the DP A$_{2}$BB$^{\prime}$O$_{6}$ (La$_{2}$NiMnO$_{6}$) is only possible in the majority spin channel,
 rather than in the minority spin channel as in 
 DP-s like SFMO~\cite{FGuinea}and SCOO~\cite{Sr2CrOsO6}} (See Appendix).

Antisite disordered regions (with B,B$^{\prime}$ interchanged) have strong antiferromagnetic superexchange 
between two nearest-neighbour B site ions, eg. half-filled Fe$^{3+}$ ions 
in case of SFMO~\cite{meDD}, or half-filled $Mn^{4+}$ ions in case of LNMO.
Hence in the disordered case, the following terms are added~\cite{FGuinea}:
\begin{eqnarray}
H_{disord}&=&t_{MM}\sum_{<ij>}\left[cos\frac{\theta_{i}}{2}cos\frac{\theta_{j}}{2}\right. \nonumber \\
          &+&\left.e^{i(\phi_{j}-\phi_{i})}sin\frac{\theta_{i}}{2}sin\frac{\theta_{j}}{2}\right] m_{i}^{\dagger}m_{j} \nonumber \\
&+&t_{NN}\sum_{<ij>\sigma}c_{N,i\sigma}^{\dagger}c_{N,j\sigma} 
          +J_{AS}\sum_{<ij>}\vec{S}_{i}\cdot\vec{S}_{j}
\label{antisite}
\end{eqnarray}
where $J_{AS}$ is an antiferromagnetic superexchange in the antisite region between two
neighbouring Mn $t_{2g}$ core spins (see Fig~\ref{exchangepath}),
 while $t_{MM}$ and $t_{NN}$ represent hopping between two neighbouring Mn 
 and two neighbouring Ni $e_{g}$ levels respectively.

\section{Ordered case: Dispersion and DOS}

In the ordered case, in the limit $|J_{H}|\rightarrow\infty$, the dispersion can be obtained analytically
from Eq~\ref{Jinfty} in the ferromagnetic phase. There are 3 $e_{g}$ bands, given by
\begin{eqnarray}
\epsilon_{1}&=&\epsilon_{Ni}-z\frac{J_{super}S}{2} \\
\epsilon_{\pm}&=&0.5\left[(\epsilon_{Mn}+\epsilon_{Ni})+\frac{(J_{H}+zJ_{super})S}{2}\right. \nonumber \\
&\pm&\left.\sqrt{\left\{(\epsilon_{Mn}-\epsilon_{Ni})+\frac{(J_{H}-zJ_{super})S}{2}\right\}^{2}+4\epsilon_{k}^{2}}\right]
\label{dispersion}
\end{eqnarray}
where $\epsilon_{k}=2t_{MN}(cosk_{x}+cosk_{y})$~\cite{Millis}. 
 
At half-filling of the Ni $e_{g}$ orbital in this single orbital model in the ferromagnetic state with all the
 core spins $\vec{S}$ pointing up, the Ni $e_{g}$ 
 band in only one spin channel (the majority spin channel) is fully filled, and so the Fermi energy lies in a gap.
 This gap can be estimated in the limit of Ni-Mn $e_{g}-e_{g}$ hopping $t_{MN}=0$, when the bands shrink to levels.
 Then $\epsilon_{1}\approx\epsilon_{Ni}-\frac{zJ_{super}S}{2}$,
$\epsilon_{+}\approx\epsilon_{Mn}+\frac{J_{H}S}{2}$, and
$\epsilon_{-}\approx\epsilon_{Ni}+\frac{zJ_{super}S}{2}$. 
$\epsilon_{1}$ and $\epsilon_{-}$ now represent the two spin-split Ni $e_{g}$ down (minority) and up (majority) levels
 respectively, while $\epsilon_{+}$ represents the energy of the up (majority) Mn $e_{g}$ orbital.
 The down (minority) Mn $e_{g}$
 orbital is shifted out to $\infty$ due to the $J_{H}\rightarrow-\infty$ limit being taken. Hence the energy gap in the
 majority spin channel between the occupied Ni orbital given by $\epsilon_{-}$ and the unoccupied Mn orbital given
 by $\epsilon_{+}$ is given by 
\begin{eqnarray}
E_{g}&\approx&\epsilon_{+}-\epsilon_{-} \nonumber\\
     &\approx& \Delta-\frac{zJ_{super}S}{2}  
\end{eqnarray}
  Hence the gap in the majority spin channel can be estimated as
  $\Delta-\frac{zJ_{super}S}{2}$. Thus, a necessary condition for the Ferro-Insulating state is $\Delta>\frac{zJ_{super}S}{2}$. If the Ni-Mn hopping is turned on, then this condition becomes more stringent:
\begin{equation}
\Delta-\frac{zJ_{super}S}{2}> 8t_{MN}
\label{ferroinscond}
\end{equation}
 However, if there was no Ni-Mn superexchange, the two Ni $e_{g}$ levels for up and down spins would have coincided,
 and the Ni $e_{g}$ band being half-filled, the system would have been metallic. In presence of the Ni-Mn
 superexchange, the Ni $e_{g}$ levels are spin-split. 
 The condition for non-overlapping of the Ni $e_{g}$ bands can be estimated as $\epsilon_{1}-\epsilon_{-}>0$, 
 which gives $-zJ_{super}S>0$. 
 This along with the inequality~\ref{ferroinscond} are the conditions for the realization of
 a Ferromagnetic Insulator (FI) ground state.
 {\em Hence the Ni-Mn superexchange,
 introduced in a model for LNMO in Eq~\ref{finiteJH} and Eq~\ref{Jinfty}, is an important and essential 
 component for obtaining the ferromagnetic insulating ground state in the ordered case}. 
Thus electron correlations
 are an essential criterion for the realization of the FI ground state in this model of LNMO.

In the case of finite $J_{H}$, there are 4 $e_{g}$ bands.
 In Fig~\ref{DOS}, the DOS is plotted for $J_{H}=-0.9eV$~\cite{Tanusri} by numerically solving the 4$\times$4 eigenvalue
 problem from Eq~\ref{finiteJH} in the ferromagnetic ground state for each $\vec{k}$ and using $\rho(E)=\frac{1}{N}\sum_{k}\delta(E-\epsilon_{k})$. 
The $t_{2g}$ bands for Mn and Ni do not appear in the DOS as the $t_{2g}$ electrons for Mn$^{4+}$ and Ni$^{2+}$ have been
 considered to be classical core spins of S=3/2 and S=0 respectively.
With the Mn core $t_{2g}$ spin considered to be in up state at all sites,
 it is found that the minority (down) spin Ni $e_{g}$ band and the majority (up) spin Mn $e_{g}$ band 
lie in between the majority (up) spin Ni $e_{g}$ and the minority (down) spin Mn $e_{g}$  bands, as in the DFT DOS 
of Ref~\cite{Tanusri}. The Fermi energy lies in between the Ni $e_{g}$ and the Mn $e_{g}$ majority spin  
 bands. Thus the ferromagnetic insulating state is explained. The separation between the Ni $e_{g}$ and Mn $e_{g}$
 bands in the minority spin channel is almost twice that in the majority spin channel, once again similar to the DFT DOS.
 The band gap in the majority spin channel ($\approx 1 eV$) is also reproduced. Thus the relative band positions are  
 roughly similar to that of the published DFT results~\cite{bandwidth}.

\section{Ordered case: Magnetism}

\begin{figure}
\includegraphics[width=8cm]{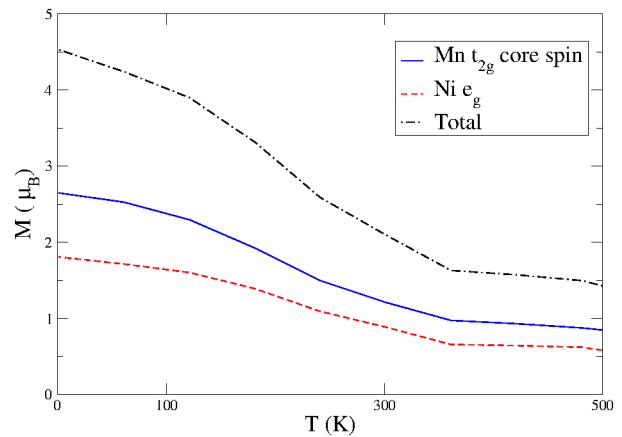} \\
\caption{(colour online) Mn $t_{2g}$ core spin, Ni $e_{g}$ electron spin and total moment as a function of temperature in the ordered case}  
\label{moment}
\end{figure}

\begin{figure}
\includegraphics[width=8cm]{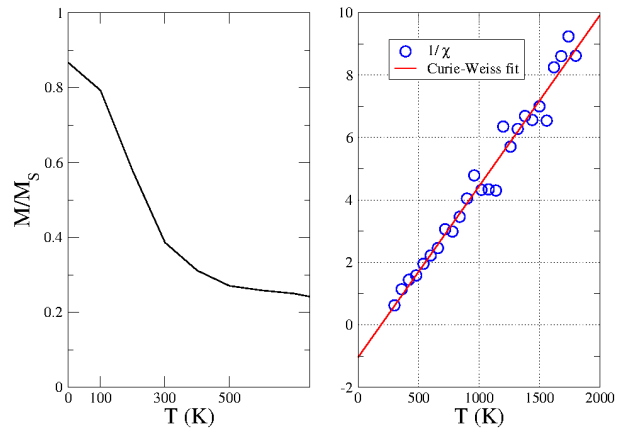} \\
\caption{(colour online) Left : $M$-$T$ plot  
 for the ordered case in 2D.
  Right: Curie Weiss fit to the reciprocal susceptibility for the ordered case.}
\label{ordered}
\end{figure}

Exact Diagonalization-Monte Carlo (ED-MC) simulations were preformed with the hamiltonian $H^{(1)}_{ord}$
  given by Eq~\ref{Jinfty},
 and a system size of 8$\times$8 in 2D, and 8$\times$8$\times$8 in 3D.
 The moments at the Nickel site, Manganese site and 
 the total moment are plotted in Fig~\ref{moment}. It is observed that
 the B and B$^{\prime}$ site moments are parallel to each other, rather than antiparallel as in Sr$_{2}$FeMoO$_{6}$. 
This is a consequence of the majority spin hybridization between B and B$^{\prime}$ site electrons in the hamiltonian of Eq~\ref{Jinfty} for LNMO, as opposed to minority spin hybridization as in SFMO~\cite{FGuinea}. This explains why LNMO
 is ferromagnetic, as opposed to SFMO, which is ferrimagnetic. This is also supported by the fact that the Nickel 
$e_{g}$ up and down spin bands do not lie within the exchange gap of the Mn $e_{g}$ bands (see DOS of 
 Fig~\ref{DOS}), as opposed to SFMO where the spin-split Mo orbitals lie within the exchange gap of Fe,
inducing a moment in Mo opposite to Fe~\cite{DDSFMO}.  The magnetization (M) versus temperature in 2D plotted in 
left pannel of Fig~\ref{ordered} shows a single transition with a $T_{c}$ around 360K, 
 while the Curie-Weiss fit of the inverse susceptibility
 gives a $T_{c}$ around 250K. The M vs T plot in 3D is shown in the left panel of Fig~\ref{mag3D}. The $T_{c}$ is 
similar (around 350K). The parameters used were: $t_{MN}$=0.125 eV,$\Delta$=1.9eV,$J_{super}$S=-7.5 meV 
(2D) and -5 meV (3D)~\cite{2d3d,Coulomb}. 
 The ordered moment reaches 90$\%$ of the maximum value
 at a temperature of about 1 K. 

\begin{figure}
\includegraphics[width=8cm]{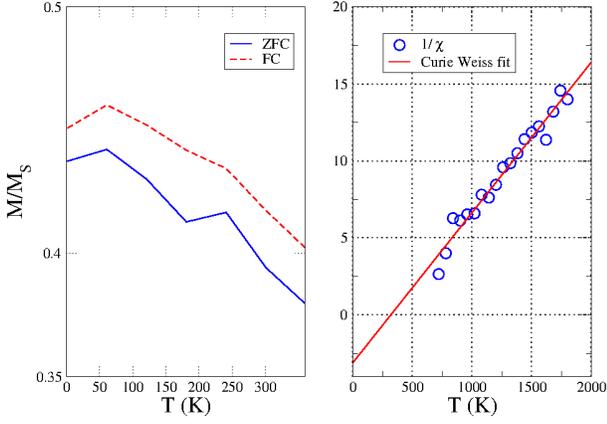} \\
\caption{(colour online) Left: ZFC-FC plots for the magnetization in the case of 25$\%$ antisite disorder in 2D.
 Right: Curie Weiss fit to the reciprocal susceptibility for the disordered case.}
\label{disordered}
\end{figure}
\begin{figure}
\includegraphics[width=8cm]{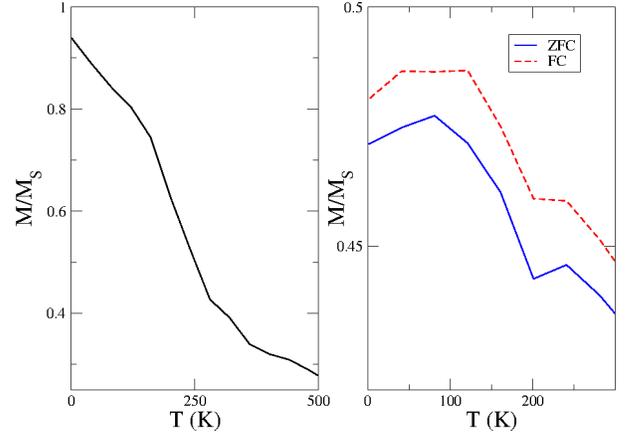} \\
\caption{(colour online) Left:  Magnetization vs temperature plot in the ordered case in 3D.
 Right: ZFC-FC plots for magnetization vs temperature in the case of 25$\%$ antisite disorder in 3D.}
\label{mag3D}
\end{figure}

The effective exchange for an effective B-site (Mn) core spin-only model can be calculated 
from the Hamiltonian of Eq~\ref{Jinfty}  by integrating out the B$^{\prime}$ (Ni) sites of the DP La$_{2}$NiMnO$_{6}$, 
 using the procedure of Self-Consistent Renormalization (SCR)~\cite{PinakiSCR,mePinaki}.
 As in Ref~\cite{Sr2CrOsO6}, if we
assume a onsite anisotropy on the Ni site then the $J_{super}$ term in Eq~\ref{Jinfty} becomes diagonal. In the case 
 of all spins lying parallel to this anisotropy axis ($\theta=0$ or $\theta=\pi$), the effective exchange for a
 Mn core-spin-only model~\cite{mePinaki,Sr2CrOsO6}
 $H=\sum J_{ij}^{eff}\sqrt{\frac{1+\vec{S}_{i}\cdot\vec{S}_{j}}{2}}$ can be 
evaluated as (considering majority spin hybridization with $J_{H}\rightarrow-\infty$ rather than minority spin 
 hybridization with $J_{H}\rightarrow\infty$ as in Ref~\cite{Sr2CrOsO6}):
\begin{eqnarray}
J_{ij}^{eff}&=&\sum_{k}\frac{1}{2}[E_{k+}n_{F}(E_{k+})+E_{k-}n_{F}(E_{k-}) \nonumber \\
&-&(\Delta-J^{\prime}_{super})n_{F}(\Delta-J^{\prime}_{super})]e^{i\vec{k}\cdot(\vec{r}_{i}-\vec{r}_{j})}
\label{effexchange}
\end{eqnarray} 
where $E_{k\pm}=\frac{(\Delta-J^{\prime}_{super})\pm\sqrt{(\Delta-J^{\prime}_{super})^{2}+4\epsilon_{k}^{2}}}{2}$, $J_{super}^{\prime}=zJ_{super}S/2$. However, since the 
 the Ni-Mn superexchange $J_{super}$  is ferromagnetic, $J^{\prime}_{super}<0$, unlike the Cr-Os superexchange $J_{2}$ as defined
 in Ref~\cite{Sr2CrOsO6} which is antiferromagnetic, $J_{2}>0$. 
  Hence the effective exchange expression becomes identical 
 to that of Ref~\cite{Sr2CrOsO6} with $|J^{\prime}_{super}|$ instead of $|J_{2}|$. 

In LNMO, only the lowest band out of the 3 bands 
 given in Eq~\ref{dispersion} is occupied: this signifies the Nickel $e_{g}$ 
majority spin band. Shifting
the energies of the 3 bands by $-\frac{zJ_{super}S}{2}$, and putting 
$\tilde{\epsilon}_{Mn}=\epsilon_{Mn}+\frac{J_{H}S}{2}=\Delta$ and $\epsilon_{Ni}=0$,
 the dispersions of the 3 shifted
 bands become:
\begin{eqnarray}
\epsilon^{\prime}_{1}&=&-zJ_{super}S \nonumber \\
\epsilon^{\prime}_{\pm}&=&0.5\left[\left(\Delta-\frac{zJ_{super}S}{2}\right)\right. \nonumber \\
&\pm&\left.\sqrt{\left(\Delta-\frac{zJ_{super}S}{2}\right)^{2}+4\epsilon_{k}^{2}}\right]
\end{eqnarray}
Hence there are 3 shifted bands centered at $\approx 0$, $-zJ_{super}S$ and $\left(\Delta-\frac{zJ_{super}S}{2}\right)$.
Out of these, in LNMO, only the lowest Ni $e_{g}$ band, signified by $\epsilon^{\prime}_{-}$, is occupied, as the 
electron filling is 1 per Ni $e_{g}$ orbital (this is a single orbital model).
Hence in the expression for effective exchange between Mn core $t_{2g}$ classical core spins
 given by Eq~\ref{effexchange},
  only the Fermi function for $E_{k-}$ is non-zero at T=0.
In the limit of small Ni-Mn hopping compared to Ni-Mn charge transfer energy and superexchange ( 
$\frac{t_{MN}^{2}}{\Delta-J_{super}^{\prime}}\rightarrow 0$), 
$E_{k-}\rightarrow\frac{1}{2}\left(\Delta-J_{super}^{\prime}\right)\left[1-\sqrt{1+\frac{4\epsilon_{k}^{2}}{\left(\Delta-
J_{super}^{\prime}\right)^{2}}}\right]
\rightarrow\frac{1}{2}\left(\Delta-J_{super}^{\prime}\right)\left[1-\left\{1+\frac{2\epsilon_{k}^{2}}{\left(\Delta-
J_{super}^{\prime}\right)^{2}}\right\}\right] 
\rightarrow\frac{-\epsilon_{k}^{2}}{\left(\Delta-
J_{super}^{\prime}\right)}$.
 Hence 
\begin{eqnarray}
J^{eff}_{ij}&\rightarrow&\frac{1}{2}\sum_{k}E_{k-}e^{i\vec{k}.(\vec{r}_{i}-\vec{r}_{j})} \nonumber \\
&\rightarrow&\sum_{k}\frac{-\epsilon_{k}^{2}}{2(\Delta-J_{super}^{\prime})}e^{i\vec{k}.(\vec{r}_{i}-\vec{r}_{j})} \nonumber \\
&\rightarrow& \frac{-h_{ij}}{2(\Delta-J_{super}^{\prime})}
\label{Jijapprox}
\end{eqnarray}
where $h_{ij}=t_{MN}^{2}(\sum_{2\hat{x}}\delta_{i+2\hat{x},j}+2\sum_{\hat{x}+\hat{y}}\delta_{i+\hat{x}+\hat{y},j}+4\delta_{ij})$is the Fourier transform of $\epsilon_{k}^{2}$ defined in Ref~\cite{mePinaki}. It involves a third neighbour
 term, a next-nearest neighbour term, and an onsite term.
 Hence the effective
 exchange between large B site classical core spins (Mn $t_{2g}$ core spins) in LNMO,
  with a filling of one electron  per Ni $e_{g}$ orbital,
 is {\em ferromagnetic}, just like that (between Cr $t_{2g}$ core spins) in SCOO~\cite{Sr2CrOsO6}, 
 which has a filling of one electron per Os orbital. Thus the {\em core spin ferromagnetism} of LNMO 
 arises from a similar interplay of double exchange
 $J_{H}$  and superexchange $J_{super}$ as in SCOO, except that these are both ferromagnetic in the former,
  while both are antiferromagnetic in the latter. Thus when the B$^{\prime}$ sites are included, these two exchanges
 produce overall ferromagnetism in LNMO, and overall ferrimagnetism in SCOO.

\begin{figure}
\includegraphics[width=8cm]{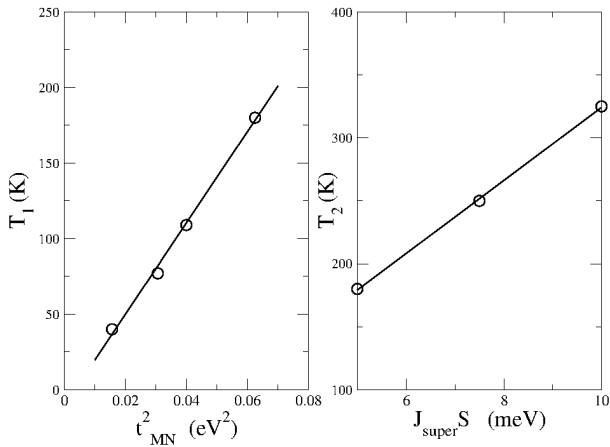} \\
\caption{Variation of the two temperatures of the kink anomalies in the magnetization with parameters, 
obtained from ED-MC simulations. 
 Left: Variation of temperature $T_{1}$ of lower temperature anomaly with $t_{MN}^{2}$, for constant $J_{super}$.
 Right: Variation of temperature $T_{2}$ of higher temperature anomaly with $J_{super}$, for constant $t_{MN}$.}
\label{T1T2}
\end{figure}
\begin{figure}
\includegraphics[width=8cm]{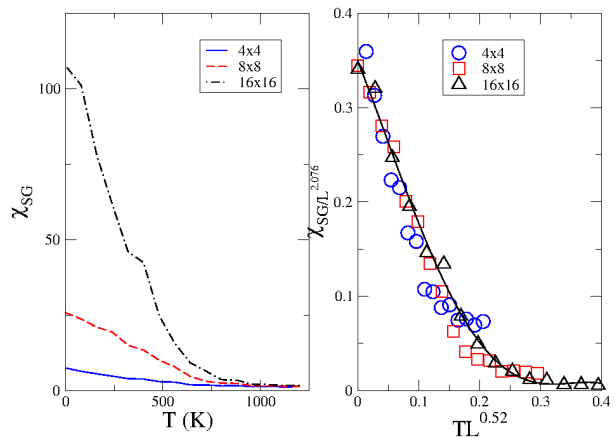} \\
\caption{(color online) Left: Spin-glass susceptibility vs temperature for the case of 25\% antisite
 disorder in 2D. Right: Data collapse for finite size scaling with $T_{SG}<0.01,\eta=-0.076,\nu=1.92$
 (T in eV)~\cite{TSG}.}
\label{spinglass2D} 
\end{figure}
\section{Disordered case: Reentrant spin-glass transition}

     ED-MC simulations with $H=H^{(1)}_{ord}+H_{disord}$ for the case of 25$\%$ random antisite disorder were
 performed with a maximum system size of 16$\times$16 in 2D, and 8$\times$8$\times$8 in 3D. The results are
    shown in Fig~\ref{disordered}, the right panel of Fig~\ref{mag3D}, and in Fig~\ref{T1T2} and Fig~\ref{spinglass2D}.
 ZFC and FC plots for the magnetization are shown in the left panel of Fig~\ref{disordered} for 2D .
 Parameters chosen are similar to the ordered case~\cite{disordparam}. The 3D results for the ZFC and FC plots
 of the magnetization are shown in the right panel of Fig~\ref{mag3D}.
 The ZFC magnetization shows a kink at around 250K corresponding to a transition to a disordered ferromagnetic state,
  followed by another kink at  around 50K, signifying the onset of a new frustrated regime, 
 where the system exhibits a tendency to become a spin-glass at low temperatures.
 This is similar to the signature of the reentrant spin-glass transition observed
 experimentally by D. Choudhury et.al.~\cite{debraj}. The ZFC-FC diverges throughout this temperature range, 
and the moment reaches only about 45-48$\%$ of its saturation value. 
The Curie Wiess fit to the high temperature susceptibility gives a $T_{c}$ of  about 320K 
 while the low temperature kink anomaly in the magnetization
  starts around 150K, and the highest value of moment is reached around 50K, indicative of the frustration in the
 system~\cite{frustration}.
 
                         In order to explore the systematics of the two kink anomalies 
in the magnetization vs temperature curve, ED-MC simulations were carried
 out for the 25$\%$ disordered systems with varying parameter sets 
(not just the parameter set obtained from DFT data of Ref~\cite{Tanusri} quoted before). 
 The results are shown in Fig~\ref{T1T2}. 
 It is observed that the temperature $T_{1}$ for the low temperature anomaly  
 varies as the square of the Mn-Ni hopping amplitude $t_{MN}$ (left panel of Fig~\ref{T1T2}),
 when $J_{super}$ is maintained constant.
 Whereas, the temperature $T_{2}$
 at which the high temperature anomaly occurs varies proportional to $J_{super}$ 
 (right panel of  Fig~\ref{T1T2}), when $t_{MN}$ is maintained constant.

 In order to confirm the spin-glass behaviour of this disordered system,
 the spin-glass susceptibility
~\cite{Fischer,BhattYoung} is plotted versus temperature in the left panel of Fig~\ref{spinglass2D}, 
in 2D for system sizes 4$\times$4, 8$\times$8 and 
16$\times$16 respectively. It is found to diverge at low temperatures, confirming that the system is indeed a spin-glass. Finite size scaling has been undertaken in the right panel of Fig~\ref{spinglass2D}. 
Upon scaling the spin-glass susceptibility as 
 $\frac{\chi_{SG}}{L^{2-\eta}}$ and plotting this versus $(T-T_{SG})L^{1/\nu}$ (where T is expressed in eV),
 the data for the 3 different system sizes are found to collapse to the same curve 
 for the choice $T_{SG}<116K, \eta=-0.076,\nu=1.92$~\cite{TSG,elphasetrans,Kawamura,YoungHeisenberg}.
 The scaling exponents $\eta$ and $\nu$ are intermediate between those for the disordered 2D Ising model~\cite{BhattYoung}
 and the disordered 3D classical Heisenberg model~\cite{YoungHeisenberg}.

As the ordering temperature for the high temperature ferromagnetic phase is $\approx$ 250-300K, which is close to the
 $T_{c}$ of the ferromagnetic phase in the ordered case, this ordering scale is clearly set by the Mn-Ni superexchange
 $zJ_{super}S$ ($\approx 300K$). 
The low temperature frustrated phase accompanied by a kink anomaly in the magnetization, is
 presumably due to the competition of the ferromagnetic Mn-Mn effective double exchange scale set by $\frac{t_{MN}^{2}}{2(\Delta-J_{super}^{\prime})}$ ($\approx 50K$, from $J^{eff}_{ij}$ in Eq~\ref{Jijapprox}),
  with the antiferromagnetic antisite Mn-Mn superexchange $J_{AS}$. 
 The presence of two ferromagnetic scales, namely due to Mn-Ni superexchange and effective Mn-Mn double exchange
 along with the antiferromagnetic antisite Mn-Mn superexchange $J_{AS}$ in LNMO presumably leads to the
 reentrant spin-glass or reentrant ferromagnetic behaviour. Such a competition between double exchange and superexchange
 leading to reentrant spin-glass behaviour have also been observed in other materials~\cite{Mathieu}. Thus,
 a rough estimate of the two temperatures related to the reentrant spin-glass transition, can be obtained as
 $T_{2}\approx zJ_{super}S$ corresponding to a transition to a high temperature superexchange dominated regime and 
 $T_{1}\approx \frac{t_{MN}^{2}}{2(\Delta-J_{super}^{\prime})}$, signifying the onset of a low temperature double
 exchange dominated regime~\cite{TSG2}.  The observed
 dependence of the two kink anomalies in the magnetization upon parameters $t_{MN}$ and $J_{super}$ 
(namely $T_{1}\propto t_{MN}^{2}$ and $T_{2}\propto J_{super}$), obtained
 from ED-MC simulations (Fig~\ref{T1T2}) as discussed before, is consistent with this picture. This establishes that
 the observed kink anomalies in the magnetization are indeed signatures of a changeover from a superexchange
 dominated to a double exchange dominated regime.

  It is to be noted that the two transitions as observed in  
 the ZFC happen in a single homogeneous phase involving Ni$^{2+}$-Mn$^{4+}$ ions and not two phases consisting of 
  Ni$^{2+}$-Mn$^{4+}$ and Ni$^{3+}$-Mn$^{3+}$ respectively, as suggested in some previous works~\cite{GoodenoughPRB}.
As is evident from Fig~\ref{moment}, the moment
 on the $e_{g}$ orbitals resides almost entirely on the Nickel, and very little on the Manganese site. 
 Thus, the 
 Nickel maintains its Ni$^{2+}$ character and the Manganese its Mn$^{4+}$ character, as reported in Ref~\cite{debraj}.
  Thus our results support the idea of a reentrant spin-glass transition within a single homogenoeus phase of
 disordered La$_{2}$NiMnO$_{6}$, as proposed in Ref~\cite{debraj}.  

\section{Conclusion}

In conclusion, a plausible explanation for the ferromagnetic insulating ground state of ordered La$_{2}$NiMnO$_{6}$ along
 with the reentrant spin-glass behaviour observed in presence of antisite disorder, is provided in a unified framework.
  The importance of the Ni-Mn superexchange
  in realizing the correlated ferro insulating state in the ordered case is established.
 Salient features of the DFT DOS are explained using this simple model Hamiltonian.
 The relevant energy scales which dictate the magnetism are identified. The underlying physics of the 
 reentrant spin-glass transition is explained in terms of a changeover from a high temperature ferromagnetic superexchange
 dominated regime to a low temperature ferromagnetic double exchange dominated regime, in competition
 with the antiferromagnetic antisite superexchange. 
  A novel mechanism of majority spin
 hybridization is proposed to explain the ferromagnetic behaviour of ordered LNMO as opposed to ferrimagnetic behaviour of
 many other DP-s like SFMO. 

\section{Appendix: Majority spin hybridization}

Let us consider a two-sublattice Kondo lattice model suitable for double perovskites,
 of the form of Eq~\ref{finiteJH}, for simplicity without the superexchange term.
\begin{eqnarray}
H_{ord}&=&\epsilon_{Ni}\sum_{i\sigma}c_{N,i\sigma}^{\dagger}c_{N,i\sigma}+\epsilon_{Mn}\sum_{i\sigma}c_{M,i\sigma}^{\dagger}c_{M,i\sigma} \nonumber \\
 &+&t_{MN}\sum_{<ij>\sigma}c_{M,i\sigma}^{\dagger}c_{N,j\sigma}
+J\sum_{i\alpha,\beta}c_{M,i\alpha}^{\dagger}\vec{\sigma}_{\alpha\beta}c_{M,i\beta}\cdot \vec{S}_{i} \nonumber \\
\end{eqnarray}
 Then the Kondo coupling term $J\sum_{i\alpha,\beta}c_{M,i\alpha}^{\dagger}\vec{\sigma}_{\alpha\beta} c_{M,i\beta}\cdot \vec{S}_{i} $ can be diagonalized by using a 
transformation of the Fermion operators $c_{M,i\uparrow}$ and $c_{M,i\downarrow}$ as follows~\cite{FGuinea}:
\begin{eqnarray}
c_{M,i\uparrow}&=&cos\frac{\theta}{2}m_{iu}+sin\frac{\theta}{2}m_{il} \nonumber\\
c_{M,i\downarrow}&=&sin\frac{\theta}{2}e^{i\phi}m_{iu}-cos\frac{\theta}{2}e^{i\phi}m_{il} 
\end{eqnarray}
For DP-s like Sr$_{2}$FeMoO$_{6}$ (SFMO), an antiferromagnetic Kondo coupling ($J>0$) 
is considered~\cite{Millis,FGuinea},
 and hence in the limit $J\rightarrow\infty$, all terms involving $m_{iu}$ operators are neglected. Then $m_{il}$ is
 set equal to spinless operator $m_{i}$. 
The hybridization terms of such a $J\rightarrow\infty$ model (as in Ref~\cite{FGuinea}), 
written in the same notation convention as followed in 
 this manuscript, is given by:
\begin{equation}
t_{MN}\sum_{<ij>} (sin\frac{\theta_{i}}{2}m^{\dagger}_{i}c_{N,j\uparrow}-e^{i\phi_{i}}cos\frac{\theta_{i}}{2}m^{\dagger}_{i}c_{N,j\downarrow}) +h.c.
\end{equation}
Obviously, if all the B site core spins $\vec{S}_{i}$ point upwards, $\theta_{i}=0,\phi_{i}=0 \forall i$, 
 whereupon the B-site spinless Fermions $m_{i}$ hybridize only with the minority down spin B$^{\prime}$ site
 electrons $c_{N,j\downarrow}$. Thus {\em minority spin hybridization} is obtained, which leads to {\em ferrimagnetism} in DP-s like SFMO, with the B$^{\prime}$
 site moment pointing opposite to the B site moment. This is because the minority B$^{\prime}$ site electrons form
 a band due to hybridization which is partially or wholly occupied, while the majority B$^{\prime}$ site electrons are
 localized, and hence remain above the Fermi energy.

On the other hand, the model for LNMO that is presented in Eq~\ref{finiteJH}  considers a ferromagnetic Kondo
 coupling, which in this case is nothing but the Hund coupling $J=J_{H}$. 
Then in the limit $J\rightarrow -\infty$, the $m_{il}$ terms are neglected, and $m_{iu}$  are set
equal to the spinless Fermion operator $m_{i}$.
The resultant model as in Eq~\ref{Jinfty}, has the following hybridization terms in the $J\rightarrow -\infty$ limit:
\begin{equation}
t_{MN}\sum_{<ij>}(cos\frac{\theta_{i}}{2}m_{i}^{\dagger}c_{N,j\uparrow}+sin\frac{\theta_{i}}{2}e^{i\phi_{i}}m_{i}^{\dagger}c_{N,j\downarrow})+h.c. 
\end{equation}
In this case, if all the B site core spins  $\vec{S}_{i}$ point upwards, 
 i.e., $\theta_{i}=0,\phi_{i}=0 \forall i$ then the B site spinless fermions $m_{i}$ hybridize only with 
 the majority spin B$^{\prime}$ site electrons $c_{N,j\uparrow}$. Hence the majority spin B$^{\prime}$ site 
 electrons form a band which in this case is fully occupied, while the minority spin B$^{\prime}$ site electrons
 are mostly localized, and remain above the Fermi energy.
Thus we get {\em majority spin hybridization} in the $J_{H}\rightarrow -\infty$ model given by Eq~\ref{Jinfty},
 leading to {\em ferromagnetism} in LNMO, with the B$^{\prime}$ moment pointing parallel to the B site moment.\\

\begin{acknowledgments}
The author acknowledges useful discussions with D. Choudhuri, H. Das, T. Saha Dasgupta and D.D. Sarma,
 and financial support through FIG 100625, and also use of the Kalam HPC cluster (DST-FIST project),
 and UNAST cluster of SN Bose National Center for Basic Sciences. 
\end{acknowledgments}


\begin{thebibliography}{99}
\bibitem{singleorb}  In LNMO, the Manganese is in Mn$^{4+}$ state and the Nickel in Ni$^{2+}$ state 
 according to Ref~\cite{debraj} and Ref~\cite{Tanusri}. The electronic configuration
   of Ni$^{2+}$ is $t_{2g}^{6}e_{g}^{2}$, while that of Mn$^{4+}$ is $t_{2g}^{3}e_{g}^{0}$.
  Thus the Nickel $e_{g}$ orbitals are
 half filled, while the Mn $e_{g}$ orbitals are empty.
 The empty $e_{g}$ orbitals of Mn$^{4+}$ being Jahn-Teller
 inactive, the electron-phonon interaction is neglected.
  Again, since the Nickel $e_{g}$ orbitals are half-filled, the orbital moment is expected to be nearly quenched.
 Thus, spin-orbit coupling is not found to affect the DFT results drastically~\cite{Tanusri} and hence is neglected.
 For these reasons, a single $e_{g}$ orbital model where the Nickel $e_{g}$ orbital
 is half filled, and the Manganese $e_{g}$ unfilled, suffices for the present study (See Fig~\ref{exchangepath}).
\bibitem{superJ1J2} Ref~\cite{Tanusri} distinguishes between the Ni-Mn $e_{g}-e_{g}$ superexchange $J(1)$
  which is FM and the Ni-Mn $e_{g}-t_{2g}$ superexchange $J(2)$ which is AFM and considers
 the net Ni-Mn superexchange $J_{net}$ 
to arise as the additive effect of the two competing interactions ($J_{net}=J(1)+J(2)$).
 This net interaction $J_{net}$ comes out to be FM and of magnitude $-5meV$. 
 However, the Ni-Mn $e_{g}-e_{g}$ superexchange cannot be considered in the present work as then the 
 Hamiltonian ceases to have spin-Fermion interactions which are coupled classical-quantum interactions, 
  and instead would have four-Fermion fully quantum mechanical interactions
  which can no longer be treated by the ED-MC
 method. The Ni-Mn $e_{g}-t_{2g}$
 superexchange can however, be treated using this method as that is between classical $t_{2g}$ core spins and quantum
 Ni $e_{g}$ Fermions. Hence this Ni-Mn $t_{2g}-e_{g}$ superexchange is used as $J_{super}$
 in the model of Eq~\ref{finiteJH} but instead considered 
 as FM, and the value of the net superexchange $J_{super}S=J_{net}=-5meV$ is used. While this is an
 approximation, it does not alter the physics greatly, as this $J_{super}$ is taken as the only Ni-Mn superexchange,
 and hence the value as well as the sign of the net Ni-Mn superexchange 
 as in the DFT paper of Ref~\cite{Tanusri} is preserved.
\bibitem{bandwidth} The bandwidths are not accurately reproduced as a single-orbital model has been considered instead
 of the two $e_{g}$ orbitals, and inter-orbital hopping has been neglected.
\bibitem{2d3d} $J_{super}S$ in 3D is 2/3 of $J_{super}S$ in 2D, so that $zJ_{super}S$ remains constant at the
 value given in Ref~\cite{Tanusri}, where $z$ is the 
 coordination number.
\bibitem{Coulomb} While the value of $J_{super}S=-5 meV$ as in
 Ref~\cite{Tanusri} correctly reproduces the ferromagnetic $T_{c}$, larger values are required to match the separation
 between the Nickel majority and minority $e_{g}$ bands in the DFT DOS . 
 The onsite Coulomb $U$ on the Nickel $e_{g}$ orbitals, neglected in this work, might have a role to play here.
\bibitem{disordparam} The parameters used for the antisite disordered part are 
 $t_{MN}=t_{MM}=t_{NN}=0.125,J_{AS}=0.025,h=0.01$. The value of $J_{AS}$ used in this work is intermediate
 between the values of nearest neighbour Heisenberg 
 Mn $t_{2g}$ core spin superexchange in rare-earth manganites (0.017 eV) as given in 
 Ref~\cite{Dagotto}, and  the value of  Fe-Fe 
 antisite superexchange for the double perovskite Sr$_{2}$FeMoO$_{6}$ (0.035 eV) considered in Ref~\cite{FGuinea}.
\bibitem{frustration} The next-nearest-neighbour B$^{\prime}$-B$^{\prime}$ (Ni-Ni) superexchange
 which introduces an additional 
 source of frustration even in the ordered case, like the Os-Os superexchange in SCOO~\cite{Mohit},
 has not been considered in this work.
\bibitem{TSG} Since the temperature scale in the data collapse plot is in eV, the error in determining $T_{SG}$
 is about 0.01 eV for 2D. Hence the only definitive statement that can be made about $T_{SG}$ is that it is less than 116K in 2D.
\bibitem{TSG2} While this temperature $T_{1}$ signifies the onset of a frustrated double exchange dominated regime, 
 when the effective Mn-Mn double exchange  
 begins to compete with the antisite Mn-Mn superexchange,
   and spin-glass correlations begin to proliferate, the actual spin-glass
 transition $T_{SG}$ may occur at lower temperatures.
\end{thebibliography}
\end{document}